\documentclass[aps,pra,reprint,twocolumn,eqsecnum,nofootinbib,longbibliography]{revtex4-2}

\usepackage{amsmath}
\usepackage{graphicx}
\usepackage{mathrsfs}
\usepackage{color}
\usepackage{hyperref}
\usepackage{lipsum}
\usepackage[procnames]{listings}
\usepackage{layouts}
\usepackage{epstopdf}
\usepackage{dsfont}
\usepackage{bbm}
\usepackage{ulem}

\hypersetup{
	colorlinks = true,
	urlcolor   = blue,
	linkcolor  = red,
	citecolor  = blue
}

\definecolor{green2}{rgb}{0,0.5,0}
\definecolor{backcolour}{rgb}{0.95,0.95,0.95}
\definecolor{grey}{rgb}{0.5,0.5,0.5}

\newcommand{\nanotrappy}{\texttt{\lowercase{nanotrappy}}}

\newcommand{\sixj}[6]{
	\begin{Bmatrix}
		#1 & #2 & #3 \\
		#4 & #5 & #6
   \end{Bmatrix}
   }

\newcommand{\rde}[2]{\langle #1||\boldsymbol{d}||#2\rangle}
\newcommand{\vect}[1]{\boldsymbol{#1}}

\begin{document}

\title{Nanotrappy: An open-source versatile package for cold-atom trapping\\ close to nanostructures}

\author{J\'{e}r\'{e}my Berroir}
\thanks{These authors contributed equally to this work.}

\author{Adrien Bouscal}
\thanks{These authors contributed equally to this work.}

\author{Alban Urvoy}

\author{Tridib Ray}

\author{Julien Laurat}
\email[email: ]{julien.laurat@sorbonne-universite.fr}

\affiliation{Laboratoire Kastler Brossel, Sorbonne Universit\'e, CNRS, ENS-Universit\'e PSL, Coll\`{e}ge de France, 4 place Jussieu, 75005 Paris, France}

\date{\today}
\begin{abstract}
Trapping cold neutral atoms in close proximity to nanostructures has raised a large interest in recent years, pushing the frontiers of cavity-QED and boosting the emergence of the waveguide-QED field of research. The design of efficient dipole trapping schemes in evanescent fields is a crucial requirement and a difficult task. Here we present an open-source Python package for calculating optical trapping potentials for neutral atoms, especially in the vicinity of nanostructures. Given field distributions and for a variety of trap configurations, \nanotrappy{} computes the three-dimensional trapping potentials as well as the trap properties, ranging from trap positions to trap frequencies and state-dependent light shifts. We demonstrate the versatility for various seminal structures in the field, e.g., optical nanofiber, alligator slow-mode photonic-crystal waveguide and microtoroid. This versatile package facilitates the systematic design of structures and provides a full characterization of trapping potentials with applications to coherent manipulation of atoms and quantum information science.

\end{abstract}

\maketitle

\section{Introduction}\label{sec:intro}

Integrating cold atoms and nanophotonic devices enables to create original light-matter interfaces. This effort has paved the way for cavity-QED platforms with unprecedented atom-field coupling via nanoscale cavities \cite{Lukin2013,Lukin2014} and for new opportunities by coupling atoms and guided light. Waveguide-QED platforms have emerged with promises not only for developing quantum information network capabilities but also as a new paradigm for creating exotic quantum phases of light and matter \cite{RMPKimble}. In this context, a number of capabilities were demonstrated using cold atoms coupled to the evanescent field of optical nanofibers \cite{Nieddu2016,Solano2017a,Nayak2007,Mitsch2014,Laurat2015,Sayrin2015,Laurat2016,Polzik2016,Solano2017,Polzik2018,Laurat2019,Kato2019,Nicchormaic2020,Nicchormaic2020a,Prasad2020}. More recently, periodic structuring into bandgap-engineered photonic-crystal waveguide has raised a large interest to strongly enhance atom-photon coupling in single pass or provide band-gap-mediated atom-atom interaction \cite{RMPKimble,Goban2014,Kimble2016,Burgers2019}.

The key ingredient for achieving strong atom-photon coupling, is the ability to trap atoms using the evanescent field of the guided modes. This is a challenging task. In a nanofiber-based platform, thanks to a featureless dispersion relation, a two-color evanescent field trap \cite{LeKien2004,Vetsch2010,Kimble2012a} is commonly used. For more complex structures, e.g., microtoroid or photonic-crystal waveguide, atom trapping via guided modes has remained a roadblock and side illumination was mostly used heretofore \cite{Lukin2013,Goban2014,Kimble2020,Arno2021}. A specific feature of such optical microtraps is the strong gradients of electric fields and polarization, which can introduce a spatially varying shift to the atomic energy levels, leading to inhomogeneous shift and fictitious magnetic fields \cite{Arno2016}. Hence, finding an adequate trapping scheme requires an extensive optimization process for each structure. Moreover, considering real multilevel atoms with hyperfine and Zeeman sublevels adds to the complexity. It will thus be convenient to have a versatile tool to optimize and characterize optical dipole traps, in particular near nanophotonic devices.

In this paper we present an open-source Python package, \nanotrappy{} \cite{Nanotrappyweb}, for computing state-dependent optical trapping potential for multilevel alkali atoms. It leverages upon the community maintained database accessible through the Alkali Rydberg Calculator~\cite{Robertson2021} to provide up-to-date computational results. The package provides the user with a programmatic API as well as customized graphical tools for trap vizualization and easy parameter adjustment. These are particularly useful in the context of trap design and in-situ optimization. Even though we put an emphasis on trapping in evanescent fields close to nanophotonic structures, it is a platform independent solution that adresses design and optimization problems faced by the much broader community trapping atoms with optical dipole potentials.

\begin{figure*}[t!]
\centerline{\includegraphics[width = 0.97\textwidth]{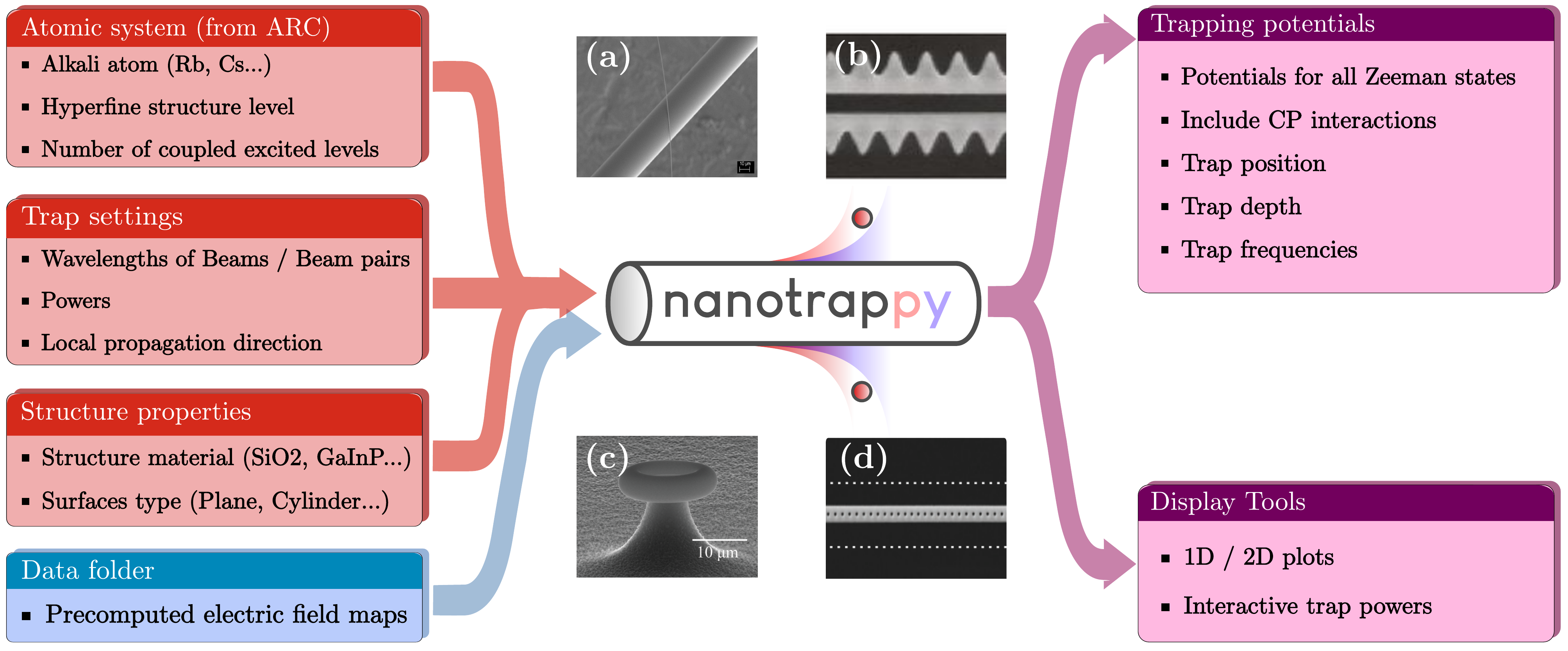}}
\caption{The \nanotrappy{} package for calculating optical trap potentials. The Python package takes multiple physical parameters as inputs (atom, trap configuration, material) as well as precomputed electric field maps, and returns the trap and its properties. A user-friendly graphical interface also enables to tune the powers of the trapping beams, making the computation easy to integrate in a structure optimization workflow. Some examples of nanostructures for which optical trapping of atoms can be simulated in \nanotrappy{}: (a) Optical nanofiber \cite{Gouraud2016}, (b)~Alligator photonic-crystal waveguide \cite{Kimble2016}, (c)~Microtoroid \cite{Alton2013}, (d)~Nanoscale optical cavity \cite{Lukin2013}.}
\label{fig:package_scheme}
\end{figure*}

The workflow of \nanotrappy{}  is illustrated in Fig. \ref{fig:package_scheme}. The package has four input classes, namely (i) the alkali atom for which the trap is to be simulated, (ii) the details of the trapping light, e.g., wavelengths, powers or propagation directions, (iii) the material of the structure for computing surface force, and (iv) the field distribution of the light around the nanophotonic structure. The latter one can be directly computed in our package for some simple structures like nanofibers, or imported, e.g. from FDTD calculations. With the above inputs, the package outputs the trapping potential for all the hyperfine Zeeman states of the ground and excited levels. It evaluates the relevant parameters of the trap such as its position, depth and frequencies along all three axes. Once the trap has been computed for a given geometry and a set of parameters, the latter can be varied in real time using the interactive GUI to optimize the trap properties.

The paper is organized as follows. In Section~\ref{sec:theory} we first provide a brief summary of the theoretical framework used for calculating the optical potentials for multilevel alkali atoms. Scalar, vector and tensor shifts are taken into account. We then describe the package architecture and operation in Section~\ref{sec:package_description}, detailing the possible inputs and outputs. The versatility of the package is demonstrated in Section~\ref{sec:application} for three examples of nanophotonic structures, namely, optical nanofiber, alligator slow-mode waveguide and microtoroid, that have been considered in recent years for trapping atoms in evanescent fields.
Section~\ref{Conclusion} concludes the paper.

\section{Theoretical framework} \label{sec:theory}

Optically trapping atoms relies on the conservative potential created by the intensity distribution of detuned laser beams~\cite{Grimm2000}, a technique now widely used for trapping cold atoms in free space.
For optically trapping atoms at sub-wavelength distance from a dielectric surface, two specificities arise.
First, the evanescent field leaking out of the dielectric material should be able to provide a stable trapping potential close to the surfaces~\cite{Balykin1991,Vetsch2010,Kimble2012a,LeKien2004}. Additionally the Casimir-Polder interaction becomes sizeable at such distances and has to be addressed~\cite{LeKien2004,Boustimi2002}.

In this section, we will briefly introduce the atom-light interaction concepts at the heart of the dipole trapping potential, with an emphasis on the influence of the Zeeman hyperfine levels. We recall the origin of scalar, vector and tensor shifts on the hyperfine manifold and their contributions to the polarizability tensor. Then, we discuss the Casimir-Polder interaction at sub-wavelength distance from the surface. Finally, all these contributions are included to compute the total trapping potential.

Note that despite the emphasis on trapping in the evanescent field of modes guided by a dielectric waveguide, the following presentation remains general and valid for any electric field distribution.

\subsection{Atom-light interaction: light shifts}\label{subseq:interaction}

We will consider here the case of a multilevel alkali atom interacting with a monochromatic optical field. To simplify the notation and describe easily the internal state of the atoms, we will adopt convenient notations, following the work of \cite{Steck2019, LeKien2013}.  We denote $|s\rangle =~|N,J,F,m_{F}\rangle$ the state of interest with bare energy $E_{s}$ and $|e_{i}\rangle = |N_{i}',J_{i}',F_{i}',m_{F,i}'\rangle$ all the states to which it is dipole-coupled with bare energies $E_{e_{i}}$. $J$ stands for the total electronic angular momenta, $F$ for the total atomic angular momenta and $m_{F}$ for the magnetic quantum number. Here we choose $z$ as the quantization axis.

The Hamiltonian of such atom-field interaction is the textbook semi-classical electric-dipole interaction~\cite{CohenBook}: ${H_{\textrm{AF}}=-\vect{d}\cdot\vect{\mathcal{E}}}$. All the trapping fields are assumed to be classical fields, which is well justified experimentally considering the powers that are typically used.

In order to study the frequency response of an electric dipole interacting with an optical electric field, we define a polarizability tensor $\alpha_{\mu\nu}$ such that the mean induced dipole moment vector becomes:

\begin{equation}
	\langle d^{(+)}_{\mu}(\omega) \rangle = \alpha_{\mu\nu}(F,m_{F};\omega)\mathcal{E}^{(+)}_{\nu} ,
\end{equation}

\noindent where $\mathcal{E}^{(+/-)}$ and $d^{(+/-)}$ denotes the positive/negative frequency terms of the electric field and dipole operator respectively, and $\mu, \nu$ are the $q$ indices of the spherical tensor components $\{-1,0,1\}$. 

Using time-dependent perturbation theory, one can derive the Kramers-Heisenberg polarizability tensor for a given state $|s\rangle$ and a given angular frequency $\omega$ for the electric field~\cite{Steckcorrection}:

\begin{equation}
  \label{eq:polarizability}
  \alpha_{\mu\nu}(s;\omega) = \sum_{i} \left(
    \frac{\langle s|d_{\nu}|e_{i}\rangle\langle e_{i}|d_{\mu}|s\rangle}
    {\hbar(\omega_{e_{i}s}-\omega)} +
    \frac{\langle e_{i}|d_{\mu}|s\rangle\langle s|d_{\nu}|e_{i}\rangle}
    {\hbar(\omega_{e_{i}s}+\omega)}
    \right).
\end{equation}

\noindent Here $d_{q}$ represents the tensor component of the dipole operator $\vect{d}$ and $\omega_{e_{i}s}=(E_{e_{i}}-E_{s})/\hbar$.

Once this polarizability tensor has been defined, we can then express the Stark shift induced by the electric field $\vect{E}$ as:

\begin{equation}
	\begin{aligned}
	\Delta E(F,m_{F};\omega) &= -\frac{1}{2}\langle \vect{d}(\omega) \rangle \cdot \vect{\mathcal{E}}	\\
	&= -\frac{1}{2} \langle \vect{d}^{(+)}+\vect{d}^{(-)}\rangle \cdot(\vect{\mathcal{E}}^{(+)}+\vect{\mathcal{E}}^{(-)}) \\
	&= -Re(\alpha_{\mu\nu})\mathcal{E}^{(-)}_{\mu}\mathcal{E}^{(+)}_{\nu} .
	\end{aligned}
\end{equation}

In order to study the effects of each component of the electric field, a usual method is to split the polarizability tensor into its irreducible parts, namely the scalar, vector and tensor polarizabilities. Such decomposition is detailed in \cite{Steck2019}. For a rank-2 tensor $\alpha_{\mu\nu}$, the decomposition reads:

\begin{equation*}
  \alpha_{\mu\nu} = \frac{1}{3}\alpha^{(0)}\delta_{\mu\nu} + \frac{1}{4}\alpha^{(1)}_{\sigma}\epsilon_{\sigma\mu\nu}+\alpha^{(2)}_{\mu\nu}
\end{equation*}

where:

\begin{equation}
  \left\{
  \begin{aligned}
  &\alpha^{(0)} = \alpha_{\mu\mu}\\
  &\alpha^{(1)}_{\sigma} = \epsilon_{\sigma\mu\nu}(\alpha_{\mu\nu}-\alpha_{\nu\mu})\\
  &\alpha^{(2)}_{\mu\nu} = \alpha_{(\mu\nu)} - \frac{1}{3}\alpha_{\sigma\sigma}\delta_{\mu\nu} .
\end{aligned}
  \right.
\end{equation}

 After adding all three contributions, we obtain:

 \begin{widetext}
 \begin{equation}\label{eq:total_shift}
	\Delta E(F,m_{F};\omega) =  -\alpha^{(0)}(F;\omega)|\mathcal{E}^{(+)}|^{2}- \alpha^{(1)}(i\vect{\mathcal{E}}^{(-)}\times\vect{\mathcal{E}^{(+)}})_{0}\frac{m_{F}}{F}- \alpha^{(2)}(F;\omega)\frac{(3|\mathcal{E}^{(+)}_{0}|^{2}-|\mathcal{E}^{(+)}|^{2})}{2}\left(\frac{3m_{F}^{2}-F(F+1)}{F(2F-1)}\right) ,
 \end{equation}

with:

\begin{equation}
\left\{
  \begin{aligned}
    \alpha^{(0)}(F;\omega) & = \sum_{F'}\frac{2\omega_{FF'}\rde{F}{F'}^{2}}{3\hbar(\omega_{FF'}^{2}-\omega^{2})} \\
    \alpha^{(1)}(F;\omega) & = \sum_{F'} (-1)^{(F+F')}\sqrt{\frac{3F(2F+1)}{2(F+1)}}\sixj{1}{1}{1}{F}{F}{F'}\frac{\omega\rde{F}{F'}^{2}}{\hbar(\omega_{FF'}^{2}-\omega^{2})}\\
    \alpha^{(2)}(F;\omega) & = \sum_{F'} (-1)^{(F+F')}\sqrt{\frac{40F(2F+1)(2F-1)}{3(F+1)(2F+3)}}\sixj{1}{1}{1}{F}{F}{F'}\frac{\omega_{FF'}\rde{F}{F'}^{2}}{\hbar(\omega_{FF'}^{2}-\omega^{2})} .
  \end{aligned}
	\right.
\end{equation}
\end{widetext}

\noindent $\alpha^{(0)}$, $\alpha^{(1)}$ and $\alpha^{(2)}$ stand for the scalar, vector and tensor polarizabilities respectively.
The typical numbers available for computations are the reduced dipole element between states of different electronic angular momentum $\rde{J}{J'}$, with the useful relation:

\begin{align}
  \rde{F}{F'} &= \rde{J}{J'} \times (-1)^{1+F'+J+I} \nonumber \\
  			 &\quad \times \sqrt{2F'+1}\sixj{J}{J'}{1}{F'}{F}{I} 
\end{align}

\noindent where $\{:::\}$ is the Wigner 6-$j$ symbol.

\begin{figure*}
  \includegraphics[width = \textwidth]{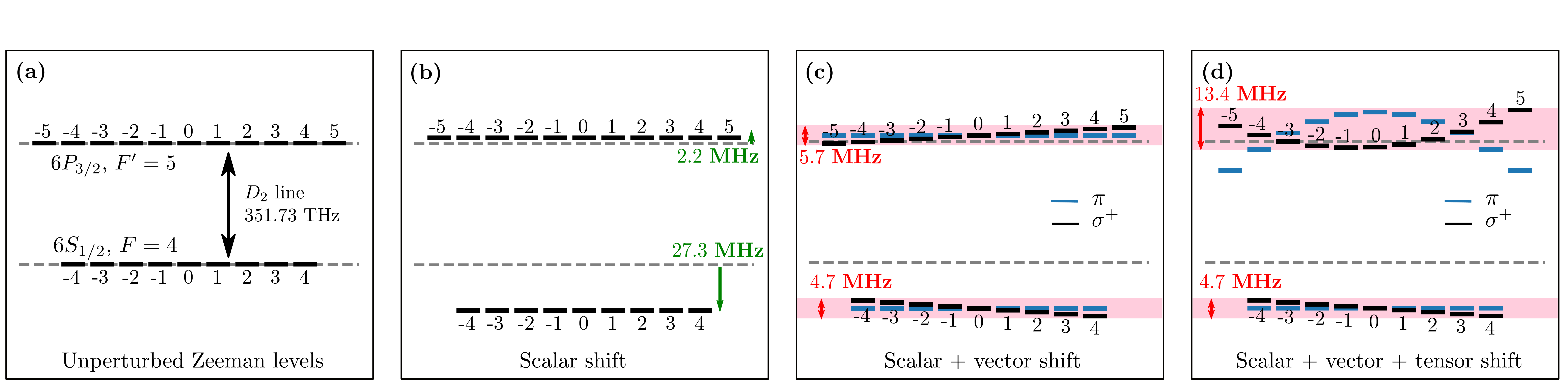}
  \caption{
    Effect of the different shift contributions induced by a $\pi$ and $\sigma^{+}$ polarized light on the $|6\mathrm{S}_{1/2},F=4\rangle$ and $|6\mathrm{P}_{3/2},F'=5\rangle$ manifolds of cesium. As a reference, numerical values for the shifts and broadenings are given for an intensity of $10$~mW/$\mu$m$^2$ at a wavelength of $1064$~nm. Vertical axis not to scale.
    (a) Unperturbed hyperfine structure.
    (b) Effect of the scalar light shift: each manifold is offset differently but irrespective of the $m_{F}$ number. This amounts to a change of the resonant transition frequency. This can be compensated using magic wavelengths, for which the shift is the same for both states: the levels are still perturbed but the transition frequency remains the same.
    (c) Effect of the scalar and vector light shifts. The linear dependence on $m_{F}$ of the vector shift is clearly visible, hence the parallel made with a fictitious magnetic field. This shift causes a broadening of the hyperfine manifold when performing spectroscopic measurements, and can be canceled by using only linearly polarized light fields.
    (d) Total shift. The tensor shift has a quadratic dependence on $m_{F}$. 
    In most cases, the unperturbed Zeemaan states $|m_{F}\rangle$ are not eigenstates of the complete Hamiltonian.
  }
  \label{fig:level_splitting}
  \end{figure*}

We now focus on the effects of the different contributions, as shown in Fig.~\ref{fig:level_splitting}.
We can see in Fig.~\ref{fig:level_splitting}(b) that the scalar shift amounts to an offset of the hyperfine manifold which depends only on the wavelength and the intensity of the light. This offset is state-dependent (between ground and excited state manifolds) and thus leads to a shift of the resonant transition.
From an experimental point of view, this shift can in some cases be suppressed by making use of the so-called magic wavelengths~\cite{Ye2008,Kimble2003}, which are available for some alkali atoms like cesium. When the trapping light is at a magic wavelength, the shift becomes state independent and thus the resonant transition frequency remains unchanged.

The form of the second term of Eq.~\eqref{eq:total_shift} highlights the different dependencies of the vector part. The main parameter is the ellipticity of the incident light, and the effect is a state-dependent shift proportional to the magnetic quantum number, which is shown in Fig.~\ref{fig:level_splitting}(c). This can be seen as the action of a fictitious magnetic field given by~\cite{LeKien2013a, Arno2016}:

\begin{equation*}
  B_{fict} = \frac{\alpha^{(1)}}{\mu_{B}g_{nJF}F}(i[\vect{\mathcal{E}}^{(-)}\times\vect{\mathcal{E}}^{(+)}]) .
\end{equation*}

\noindent This shift can be canceled by using linearly polarized light, for which the cross product vanishes and thus the fictitious magnetic field as well. In practice around nanophotonic waveguides, or in the tight-focusing regime, this is non trivial because of the strong longitudinal component of the electric field that typically introduces an ellipticity~\cite{VanMechelen2016}. In such settings counterpropagating beams have been used~\cite{Kimble2012,Laurat2019,Kien2005} to cancel out this longitudinal component.

The tensor part indicated by the third term is the most difficult to cancel in practice. We will not detail the mathematical description of this contribution and only make practical statements. Most noticeably, it vanishes for $J = 0$ and $J = 1/2$ states due to the dependence on $J$ of the tensor polarizability. However, it is not the case for excited $J=3/2$ states, which will then experience a significant tensor shift. Regarding the electric field dependence, for pure $\pi$ or $\sigma_{\pm}$ polarizations, the tensor shift part of the Hamiltonian is diagonal in the unperturbed hyperfine basis. It leaves the eigenvectors unchanged while adding a $m_{F}$-dependent shift proportional to $m_{F}^{2}$, as seen in Fig.~\ref{fig:level_splitting}(d). For arbitrary polarizations however, the tensor shift part of the Hamiltonian is non-diagonal and thus introduces coupling terms between hyperfine states. This situation is more complex and best solved numerically.

\subsection{Casimir-Polder interaction}\label{subseq:casimir}
In order to have a full description of the potential seen by the atoms close to a surface, we need to add the van der Waals interaction. These so-called Casimir-Polder (CP) interactions \cite{Casimir1948,Reynaud2017} are usually complex to calculate as they depend on the material, the atoms and the geometry of the structure.
We restrict ourselves to the first order of the potential for an infinite plane $U_{\textrm{CP}} = C_3/d^3$ \cite{Johnson2004}, where $d$ is the distance to the surface. 
This approximation puts an upper bound on the Casimir Polder effect. It is a good approximation at very low distances, and even though the curvature of the surface can lead to a 40\% error at the position of the trap minimum (in the case of a nanofiber for example), the effect remains negligible compared to the light induced potentials \cite{LeKien2004}.
Thus, we do not incorporate this correction in our calculation of the trapping potential.

The formula used to compute the $C_{3}$ coefficient for a given atom and material is taken from Ref.~\cite{Caride2005}:

\begin{equation}
  C_{3} \approx \frac{\hbar}{4\pi}\int_{0}^{+\infty} \alpha(i\xi)\frac{\epsilon(i\xi)-1}{\epsilon(i\xi)+1} \,d\xi
\end{equation}

\noindent where $\alpha(i\xi)$ is the polarizability of the trapped atom  extended over the imaginary axis, and $\epsilon(i\xi)$ is the dielectric permitivity of the material, also computed at imaginary frequencies.

\subsection{Computation of the trap}\label{subseq:trap_computation}
Building on the formalisms needed to describe both atom-light and atom-surface interactions, we can combine them to compute the full trapping potential created by arbitrary optical fields close to a surface.
In order to compute this trapping potential as well as the level mixing induced by the interactions, it is convenient to define an effective Hamiltonian for the system, which can be eventually diagonalized. For the Stark shift defined in Eq.~\eqref{eq:total_shift} this effective Hamiltonian is given by:

\begin{equation}
  \begin{aligned}
	H_\mathrm{stark} = & -\alpha^{(0)}(F;\omega)|\mathcal{E}^{(+)}|^{2}\\
   & - \alpha^{(1)}(F;\omega)(i\vect{\mathcal{E}}^{(-)}\times\vect{\mathcal{E}}^{(+)})_{0}\frac{F_{0}}{F}\\
   & - \alpha^{(2)}(F;\omega)\frac{(3|\mathcal{E}^{(+)}_{0}|^{2}-|\mathcal{E}^{(+)}|^{2})}{2}\left(\frac{3F_{0}^{2}-\vect{F}^{2}}{F(2F-1)}\right) .
  \end{aligned}
\end{equation}

Writing the CP Hamiltonian as $H_{\textrm{CP}} = U_{\textrm{CP}}\hat{\mathds{1}}$, and using perturbation theory, the total shift for a level $|N,J,F,m_{F}\rangle$ is then given by:

\begin{equation}
  \begin{aligned}
  \Delta &E_{|N,J,F,m_{F}\rangle}(\omega) = \\
  &\langle N,J,F,m_{F}|H_\mathrm{stark}(F;\omega)+H_{\textrm{CP}}|N,J,F,m_{F}\rangle .
  \end{aligned}
\end{equation}

It is important to note that in general, the Hamiltonian is not diagonal in the $|N,J,F,m_{F}\rangle$ basis because of the vector and tensor terms. Therefore it is necessary to diagonalize it in order to obtain both the correct eigenvalues and eigenvectors. While for low power, $F$ is still a good quantum number, it is not the case anymore for $m_{F}$ and the interaction gives rise to level mixing inside the magnetic Zeeman manifold. 

After introducing the theoretical framework enabling to calculate trapping potentials and state dependent light shifts, we will now present its implementation in \nanotrappy{}.

\begin{figure*}
  \includegraphics[width=\textwidth]{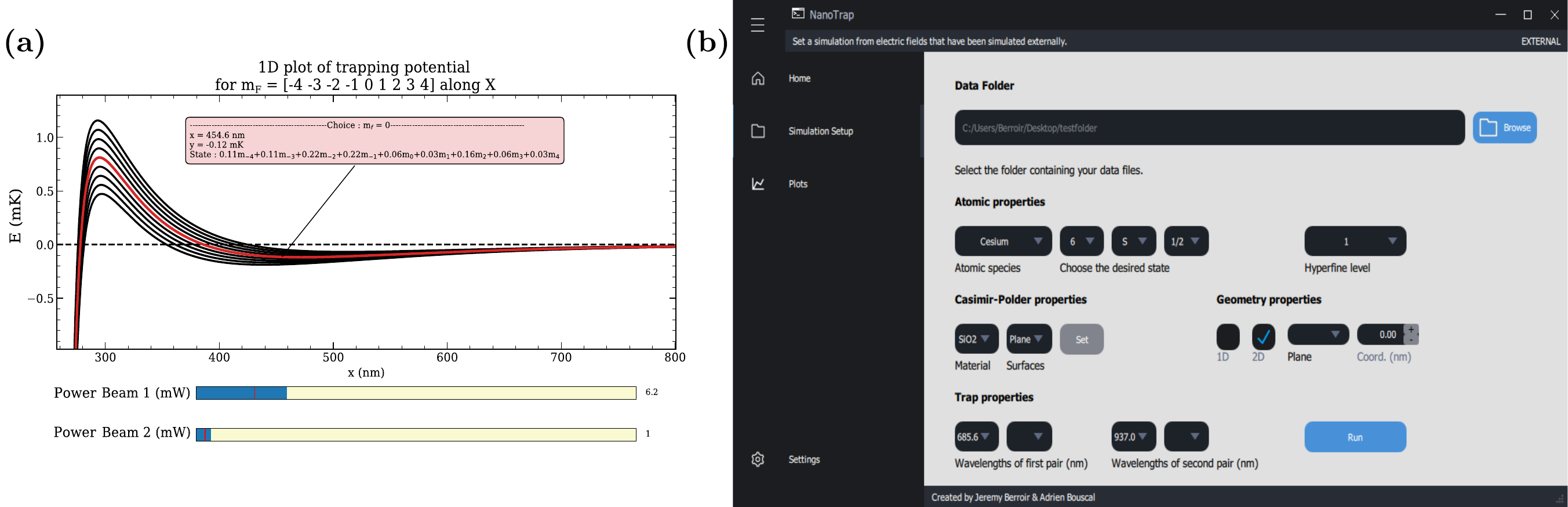}
  \caption{Screenshots illustrating the interactivity of \nanotrappy{}. (a) Interactive 1D plot with sliders (blue bars) controlling the powers of the red-detuned and blue-detuned trapping beams, here for a nanofiber configuration as described in Section~\ref{subseq:nanofiber}. The new eigenlevels can also be selected to see their decomposition on the unperturbed basis. The same controls are given for 2D plots. (b) Screenshot of the Graphical User Interface (GUI) provided as an .exe file with the package. It offers the same functionalities through dropdown menus and allows to optimize the trapping scheme without having to interact with Python code.}
  \label{fig:interactive}
\end{figure*}

\section{Package overview}\label{sec:package_description}

\nanotrappy{} is a Python package that computes the trapping potentials induced by laser beams, with an emphasis on modes guided inside nanostructures. It has been tested on Python 3.7+ distributions. \nanotrappy{} is programmed for being efficiently included in the optimization workflow of nanophotonic structure design. We take advantage of the object-oriented programming style of Python in order to provide the user with a simple and accessible API.
This allows for efficient tackling of the computation of such traps, taking multi-levels into account and using up-to-date and synchronized data referenced through ARC. 

Note that the package is not a field solver. Instead, based on a pre-computed electric field (done using any third-party solver), \nanotrappy{} interfaces this electric field distribution with an atomic system given physical parameters. A range of free and licensed field solvers already exist and specific methods maybe more suited to each use case. Thus the choice is left to the user and the package is made to be agnostic to the computational method of the optical fields.
Performance-wise, even if the limiting factor of such optimization workflows is often the actual computation of the fields, an effort has been made to make this package efficient, and to provide a parallelizable option that allows to split the computation on multiple CPU cores if needed.
In the following, we introduce the structure of the code through the base classes provided.

\subsection{Atomic system}\label{subseq:atomic_system}

The first class defined is the \texttt{atomicsystem} class. It is based on the Alkali-Rydberg-Calculator (ARC) package~\cite{Robertson2021}, which allows to select any alkali atom in the database, as well as up-to-date spectroscopic data such as transition frequencies or dipole matrix elements.
Building an \texttt{atomicsystem} amounts to selecting such an atom, as well as a state defined by the $N,L,J,F$ quantum numbers.

As \nanotrappy{} is closely linked with current experiments, it incorporates features that have proven to be crucial for the development of such systems. Among those, the issues linked with inhomogeneous broadening of an optical transition due to the Zeeman dependent nature of the light shifts call for calculations of such light shifts down to the hyperfine structure level. The polarizability of any given hyperfine state is thus computed by \nanotrappy{}, as well as $C_{3}$ coefficient of the Casimir-Polder atom-surface interaction.

\subsection{Beams and Trap}

Specific classes are used to describe the trapping light.
For each trapping beam, a \texttt{Beam} instance is created, based on a wavelength, a power and a folder containing formatted pre-computed electric fields~\cite{Formatting}.
The package will then check if the electric fields are available at that wavelength in the specified folder and select the relevant data. Counterprogating beams geometries can also be created with the \texttt{BeamPair} class.
Once the beams are created, they are bundled into the \texttt{TrapBeam} class together with a local propagation axis.

\subsection{Materials and Surfaces}
To handle CP interactions, a \texttt{Surface} class is available, as well as two main subclasses \texttt{Plane} and \texttt{Cylinder}. This allows to handle most practical cases and gives a first order approximation of these interactions, which is very often enough to get an accurate result. The CP potentials are computed for every atomic level separately.

The \texttt{Material} class comes with pre-implemented subclasses of materials (air, SiO$_{2}$, SiN, GaInP...) and can be easily extended to add other materials.

\subsection{Running the simulation}
Once all these physical parameters have been defined in the respective classes, they are bundled together in a \texttt{Simulation} class that realizes the actual computation.
Along the way, the parameters are saved as JSON files and the results as .npy numeric tables. Conveniently, a check is performed before any simulation whether these particular parameters have already been used, in which case the previously computed data is used, thereby avoiding unnecessary computation.

\subsection{Interactivity and optimization}

\begin{figure}[b!]
  \includegraphics[width = 0.97\columnwidth]{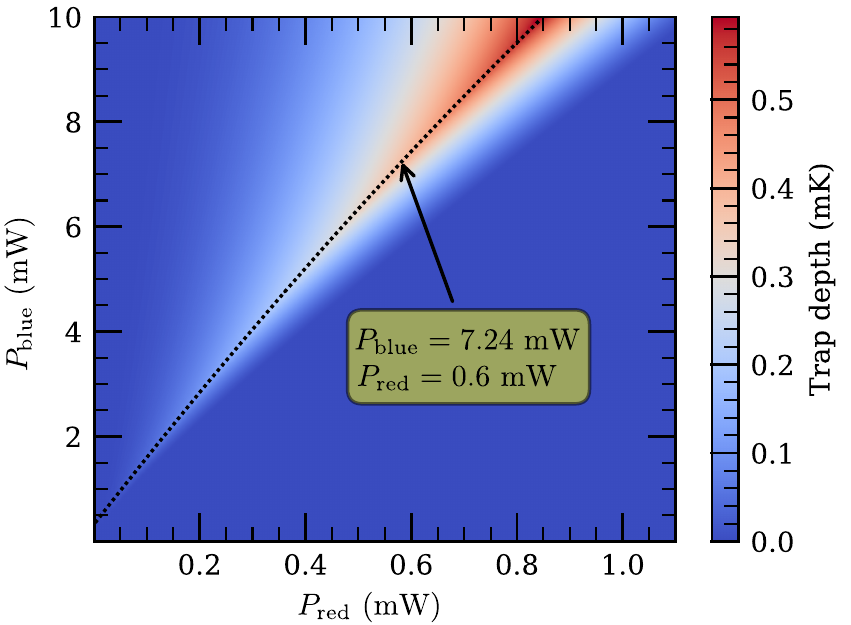}
  \caption{Automatic optimization of the trap depth for a 250~nm radius nanofiber, as in Section~\ref{subseq:nanofiber}. The powers of the 1064~nm red-detuned beams ($P_{1}$) and 780~nm blue-detuned beam ($P_{2}$) are varied and the trap depth is calculated to generate a 2D map. The dotted line follows the local trap depth maximum. Unstable trapping configurations are displayed with a trap depth of 0~mK.}
  \label{fig:optimization}
\end{figure}

The \texttt{Vizualizer} class is a core class for displaying the results, with additional features for easy optimization of the trap.
Once the simulation has run, the vizualizer will display the results along with interactive sliders that allow to control the power of each individual beam, as seen on Fig.~\ref{fig:interactive}(a). As mentioned in Section \ref{subseq:atomic_system}, it is possible to display the trapping potential for all Zeeman sublevels inside a given hyperfine state. An interactive tool allows to select a chosen Zeeman state and display additional information such as the decomposition of this new eigenstate in the basis of the unperturbed ones. If a stable trap exists, the trapping position and frequencies are also displayed. This enables to quickly and conveniently scan the powers of the beams in order to check whether the desired value for these parameters are accessible or if the structure design needs to be improved. Moreover, automatic optimization is available: the powers of the beams can be scanned to optimize a chosen parameter (either the trap depth, trap frequency or the trap position) given electric field distributions, as displayed in Fig.~\ref{fig:optimization} for a nanofiber configuration as described below in Section~\ref{subseq:nanofiber}.

A standalone GUI application is also made available with the same functionalities, and shown in Fig.~\ref{fig:interactive}(b).

\section{Package application} \label{sec:application}

In this section we show the versatility of \nanotrappy{} by computing the trapping potentials for atoms around three well known nanotructures. We benchmark \nanotrappy{}'s results against published literature to demonstrate the accuracy of the package.

\subsection{Nanofibers and uniform waveguides} \label{subseq:nanofiber}

\begin{figure*}
  \includegraphics[width = 0.95\textwidth]{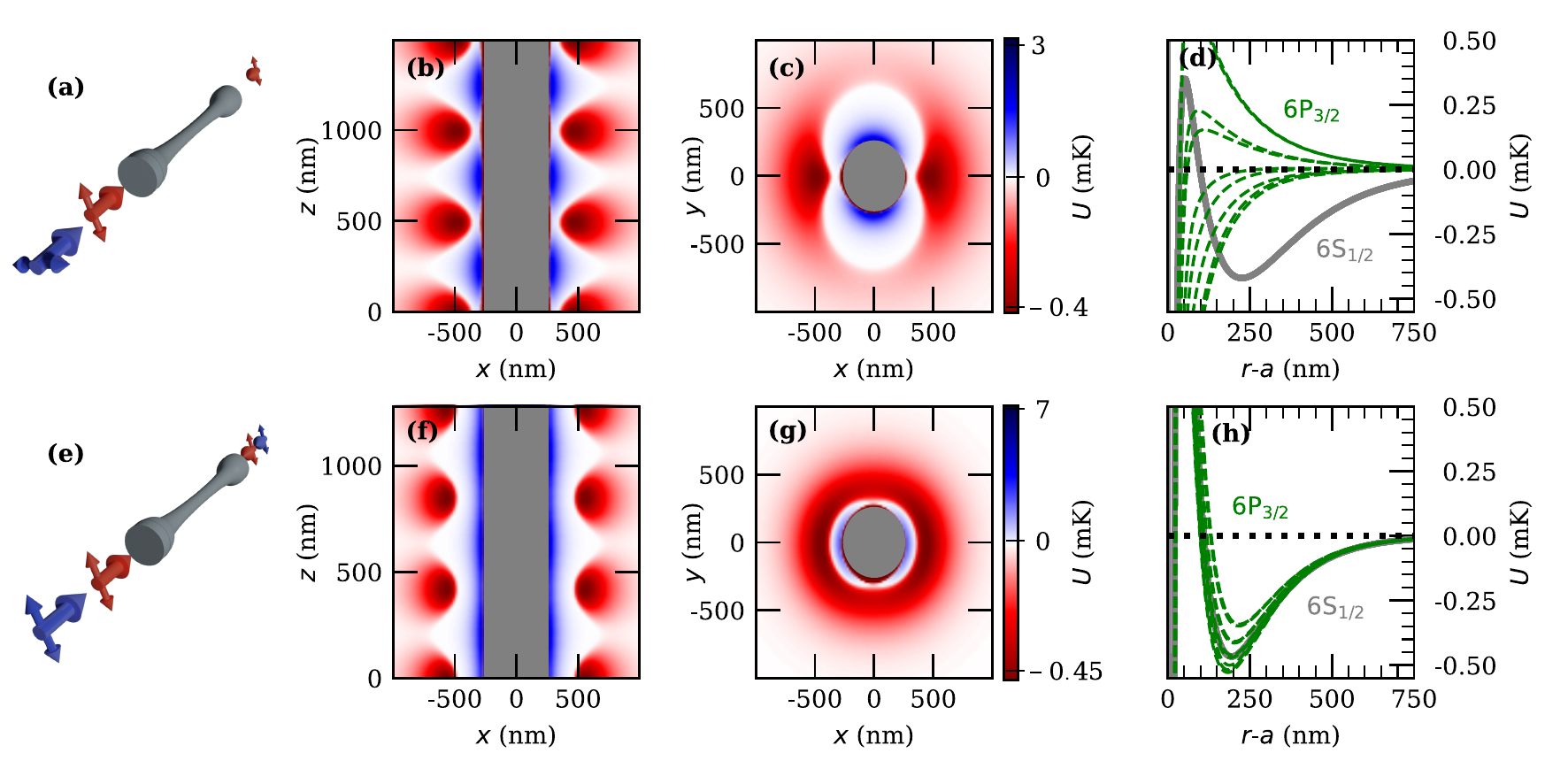}
  \caption{Two configurations of optical trapping around a nanofiber.
(a) Uncompensated nanofiber trap configuration with only one blue-detuned beam, and crossed polarisations.
(b) 2D potential $U$ ($U<0$ for stable traps) in the $xz$ plane, with same parameters as in Ref.~\cite{Vetsch2010}. Trapping sites are periodically placed with distance $\lambda_{\textrm{red}}/2$ because of the red standing wave. Stable traps with depth of around 0.4~mK are achieved.
(c) 2D potential in the $xy$ plane at the $z$-position of a trap.
(d) Radial dependence of the trapping potential of the ground ($6$S$_{1/2}$) and excited ($6$P$_{3/2}$) states along the $x$ axis. The splitting of the $m_F$ states in the ground state is not visible at this scale. The trap minimum is located at around 220~nm from the surface but the atoms in the excited state are not trapped.
(e-h) Same plots for the state-insensitive, compensated configuration (see text) with parameters from Ref.~\cite{Kimble2012}. (g) Azimuthal trapping is less efficient in this configuration but (h) a stable trap for excited atoms, with low inhomogeneous broadening is obtained.}
\label{fig:nanofiber}
\end{figure*}

Optical nanofibers have been largely used for atom-nanophotonics interfaces. Relatively simple fabrication technique of subwavelength diameter, low-loss silica nanofiber \cite{Tong2003} and easy integrability with cold atoms makes it a popular choice.
Early works involved an optical nanofiber embedded in an ensemble of atoms in a magneto-optical trap (MOT)~\cite{Nayak2007,Nayak2009}. The first proposal on trapping atoms using  the evanescent field of an optical nanofiber was by balancing the attractive gradient force of the evanescent component of a red-detuned field with the centrifugal force when the fiber diameter is about half the wavelength of the trapping light \cite{Balykin2004}. Later it was proposed that the attractive force can be counter-balanced by the evanescent component of a blue-detuned field propagating in the same nanofiber, giving birth to the seminal two-color evanescent field trap~\cite{LeKien2004,Vetsch2010}. A state insensitive, compensated trap was proposed to suppress the inhomogeneous light shifts and demonstrated for cesium atoms~\cite{Kien2005,Kimble2012a,Kimble2012}.\\

In \nanotrappy{}, both electric and magnetic fields of the guided modes of a nanofiber can be analytically computed thanks to the fiber eigenvalue equation~\cite{Snyder2012}, given a radius and a refractive index for the dielectric. This calculation is implemented in \nanotrappy{}, so that for this simple structure the package can be used for computing both the electric field distributions and the trapping potentials.

We use \nanotrappy{} to compute the trapping potentials for an uncompensated trap and a compensated one, and compare them to published literature \cite{Vetsch2010,Kimble2012}. The results are shown in Fig. \ref{fig:nanofiber}. In both cases the goal is to compute the characteristics of the traps for ground and excited state of cesium atoms around a SiO$_2$ nanofiber with 250~nm radius. The differences between the schemes come from the wavelengths, powers, polarization and number of beams used.

Figure \ref{fig:nanofiber}(a) shows the configuration of the uncompensated trap.
The parameters are chosen as per \cite{Vetsch2010}.
A pair of red-detuned, counterpropagating beams at 1064~nm and a single, blue-detuned beam at 780~nm are used to create the trapping potential.
The red- and blue-detuned beams have orthogonal linear polarization.
The total powers used are  $P_\mathrm{red}$ = 2 $\times$ 2.2 mW and  $P_\mathrm{blue}$~= 25~mW respectively.
The obtained results are shown in Fig.~\ref{fig:nanofiber}(b-d).
A 1D array of evenly spaced traps along the nanofiber with depth 0.4~mK is achieved at around 195 nm from the surface.
The corresponding trap frequencies are 355~kHz, 71~kHz, 355~kHz along $r$, $\theta$ and $z$ respectively.
We note that, in such a trap only the ground state cesium atoms are trapped, the excited $6\mathrm{P}_{3/2}$ states experience a repulsive potential as shown in \ref{fig:nanofiber}(d).
The results are in excellent agreement with \cite{Vetsch2010, VetschPhD2010}.

For the compensated trap, as shown in Fig.~\ref{fig:nanofiber}(e), the scheme and parameters are chosen as per \cite{Kimble2012}.
In this configuration, a second, counterpropagating blue beam is used in order to reduce the vector shift as much as possible.
The powers are $P_\mathrm{blue}$ = 2 $\times$ 16 mW and ${P_\mathrm{red} = 2 \times 0.95 \textrm{mW}}$.
The results are shown in Fig.~\ref{fig:nanofiber}(e-h).
Stable traps are obtained for both the ground and excited levels.
The computation yields a trap at 190~nm from the surface with depth $\sim 0.5$~mK for the ground state and 0.3 to 0.6~mK for the excited state. The results are also in excellent agreement with \cite{Kimble2012, GobanPhD2015}.

We now use this well-known example of a trap around a nanofiber to illustrate step-by-step how to compute dipole traps with \nanotrappy{}. This sample code, only a few lines long, can be easily adapted to any structure and alkali atom of interest.

\lstset{language=Python,
    backgroundcolor=\color{backcolour},
    basicstyle=\ttfamily\small,
    keywordstyle=\color{blue},
    commentstyle=\color{grey},
    stringstyle=\color{green2},
    linewidth=7.8cm,
    xleftmargin=-0.8cm,
    showstringspaces=false,
    identifierstyle=\color{black},
    procnamekeys={def,class},
    breaklines=true,
    breakatwhitespace=true,
    postbreak=\raisebox{0ex}[0ex][0ex]{\ensuremath{\hookrightarrow}}}

\begin{enumerate}
    \item First, an atomic system has to be specified. This part makes use of the ARC package~\cite{Robertson2021}, hence all alkali atoms can be used. The other parameters of the \texttt{atomicsystem} define the hyperfine level of the ground state considered for trapping, here ground state $6\mathrm{S}_{1/2}$ cesium with $F=4$.
    \begin{lstlisting}
    #Definition of the atomic system
    syst = nt.atomicsystem(Caesium(), nt.atomiclevel(6,S,1/2),f = 4)
    \end{lstlisting}

    \item The trapping scheme then has to be defined: number of beams, wavelengths, counterpropagating or not... The wavelength is of utmost importance as the package will look for the spatial mode corresponding to this wavelength in the data folder.
    \begin{lstlisting}
    import nanotrappy as nt
    #Definition of the beams used for trapping
    blue_beam = nt.Beam(780e-9, direction = "f", power = 25*mW)
    red_standing_wave = nt.BeamPair(1064e-9, power1 = 2.2*mW, 1064e-9, power2 = 2.2*mW)
    trap = nt.Trap_beams(blue_beam, red_standing_wave)
    \end{lstlisting}

    \item (Optional) The structure around which the atoms are trapped can also be defined. This is necessary for including the CP potential $U_\textrm{CP}$ (see Sec.~\ref{sec:theory}).
Infinite planes and cylinders are already implemented in the package using $U_{\textrm{CP}} = C_3/d^3$, as well as many materials. If not specified, no surface is taken into account.
    \begin{lstlisting}
    #Adding a surface for CP interactions
    surface = nt.Cylinder(radius = 250e-9, axis = AxisZ(coordinates = (0,0)))
    \end{lstlisting}
		
		\item The \texttt{simulation} object that will store the results of the calculations is created, taking as arguments all the previous objects, plus the data folder and the structure material.
    \begin{lstlisting}
    #Create the simulation object that will store the results
    Simul = nt.Simulation(syst,Nm.SiO2(), trap,datafolder,surface)
    Simul.geometry = PlaneXZ(normal_coord = 0)
    \end{lstlisting}
		
		\item Running the simulation boils down to one line of code.
    \begin{lstlisting}
    trap2D = Simul.compute()
    \end{lstlisting}
		For the $500 \times 500$ grid of Fig.~\ref{fig:nanofiber} this evaluation takes less than a minute on a standard office computer. 
		
		\item A \texttt{vizualizer} object is then created to display the results and manipulate the optical powers. After the initial computation, the trap powers can be varied, and the changes appear in real time.
    \begin{lstlisting}
    viz = nt.Vizualizer(Simul, trapping_axis="Y")
    fig, ax, slider_ax = viz.plot_trap()
    \end{lstlisting}
		\end{enumerate}

\subsection{Photonic-crystal waveguides : the APCW}\label{subseq:photonic_crystal}

\begin{figure*}
  \includegraphics[width = 0.95\textwidth]{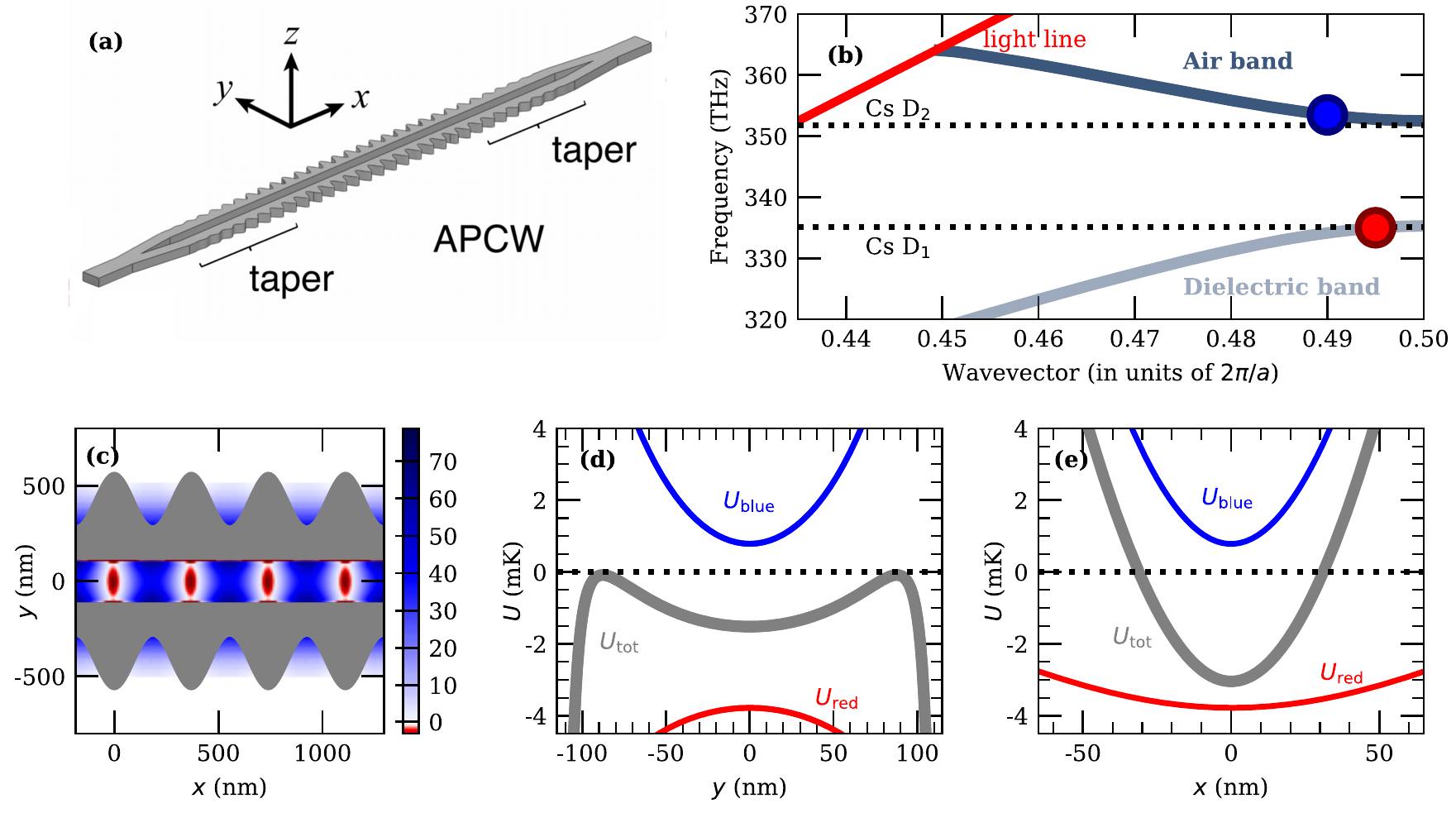}
\caption{Stable trapping of atoms inside the gap of the Caltech alligator photonic-crystal waveguide (APCW).
(a) Schematic of the full APCW, with taper regions, extracted from Ref.~\cite{Goban2015}.
(b) Optimized band structure of the APCW with band edges aligned to D$_1$ and D$_2$ lines. Blue- and red-detuned modes used for trapping are shown with the corresponding color circles. The light line for a suspended waveguide in vacuum is shown in red.
(c) 2D total trapping potential with superimposed structure. A periodic stable trap with depth of more than 3 mK are achievable with powers of 230 $\mu$W for the blue beam and 3 $\mu$K for the red one.
(d) Calculated trapping potential along the $y$ axis. The grey curve corresponds to the total potential $U_{\mathrm{tot}} = U_{\mathrm{red}} + U_{\mathrm{blue}} + U_{\mathrm{CP}}$. Blue- and red-detuned beam contributions are plotted separately for comparison.
(e) Trap along the $x$ axis. The trapping frequency in this direction is large with: $\omega_{x} = 2\pi \times 3$~MHz.}
	\label{fig:trap_APCW}
\end{figure*}

We now consider a second example involving trapping atoms around a slow-mode photonic-crystal waveguide. Such a platform with atoms trapped by evanescent modes is still to be experimentally demonstrated, but several theoretical proposals for trapping have recently emerged.
A one-color dipole trap for trapping Cesium atoms was first proposed~\cite{Yu2014,Hung2013}, using only a single blue-detuned laser from the D2 transition. But for all photonic crystals considered, this scheme leads to small trap depths of a few tens of $\mu$K.

To overcome this difficulty without increasing the powers of the trapping beams, generally limited by the power handling of such devices, a two-color dipole trap was proposed for the Caltech alligator photonic crystal waveguide (APCW)~\cite{Burgers2019}, following the ideas implemented with nanofibers. We compute the trapping potential for this APCW with \nanotrappy{} and compare them to \cite{Luan2020}. The results are presented in Fig.~\ref{fig:trap_APCW}.\\

Figure~\ref{fig:trap_APCW}(a) shows the aforementioned waveguide and Fig.~\ref{fig:trap_APCW}(b) its band structure. The parameters of the device are the same as in~\cite{Burgers2019}: the period is 370 nm, the gap is 240 nm wide, the edge modulation is 140 nm and the refractive index is 2 (for SiN). With these numbers the air and dielectric bands are aligned to the D2 and D1 lines of cesium, respectively.

Figure~\ref{fig:trap_APCW}(c) shows a trapping potential in two dimensions. The parameters for the trap are chosen as in~\cite{Luan2020}. A beam red-detuned from the D$_1$ line of Cs at 895~nm ($\delta = 2\pi \times 1700$~GHz) and a beam blue-detuned from the D$_2$ line at 848.1 nm ($\delta = 2\pi \times -130$~GHz) are used to create the trapping potential. The total powers are ${P_\mathrm{blue} = 230~\mu}$W and ${P_\mathrm{red} = 3~\mu}$W.

As shown in Fig.~\ref{fig:trap_APCW}(d-e), a stable trap in the $x$ and $y$ directions is obtained. There is also trapping in the $z$ direction, although with less strength. The trapping sites are positioned in the center of the gap, with a trap depth of 3~mK. The trapping frequencies are $\omega_{y} = 2\pi \times 1.1$~MHz, $\omega_{x} = 2\pi \times 3$~MHz and $\omega_{z} = 2\pi \times 570$~kHz. Confinement on the propagation direction is therefore very strong.
The values computed with \nanotrappy{} are in very good agreement with \cite{Luan2020}. The slight differences come mostly from numerics in the electric field simulations.

\subsection{Microtoroid}\label{subseq:microtoroid}

\nanotrappy{} is a versatile package as it can also be used for structures that are not waveguides. We demonstrate this here by studying the trapping of atoms near a microtoroid resonator.

One of the earliest proposal of trapping atoms with the evanescent field of a microstructure was to use the whispering gallery mode (WGM) of a microsphere \cite{Mabuchi1994}. The high Q factor and small mode volume of such resonators \cite{Kimble1998} can yield single-atom strong coupling on average \cite{Kimble1998a}, even with hot atomic vapors. Later, a toroidal microcavity was proposed as ultrahigh-Q microresonator for cavity QED \cite{Kimble2005}. It shows an even smaller mode volume and increased tunability arising from the added degree of freedom associated with the principal and minor diameters of the microtoroid. Experimentally, strong interaction with a WGM of a microtoroid was demonstrated with free-falling cesium atoms \cite{Kimble2006,Dayan2008}. Following these first demonstrations, schemes for trapping Cesium atoms in the evanescent field of such microresonators were proposed \cite{Stern2011, Alton2011, Alton2013}.

\begin{figure}
    \includegraphics[width = 0.97\columnwidth]{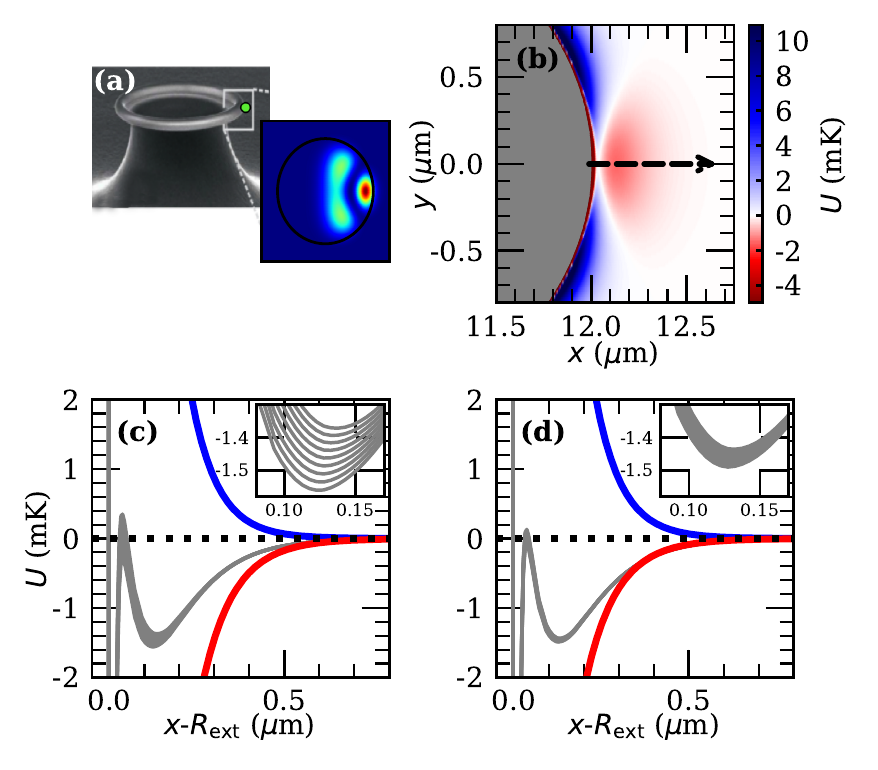}
	\caption{Two-color scheme for trapping atoms on the symmetry plane of a microtoroid resonator. (a) SEM image of a fabricated SiO$_2$ microtoroid extracted from \cite{Kimble2005}. Inset : 2D intensity profile of the red-detuned higher-order mode used for trapping.
	(b) 2D potential on the outer edge of the structure. Wavelengths of 900~nm and 850~nm are used for the red- and blue- detuned beams. The line cuts (c) and (d) are taken along the dashed arrow.
	(c) Trapping potential along $y$, at $z = 0$ and with CP potential included. The inset shows a zoom at the position of the minimum to highlight the splitting of the $m_F$ states due mostly to the vector shift.
	(d) Trapping potential along $x$ with a counterpropagating red-detuned beam. The inset shows a reduced splitting compared to (c) due to cancellation of the vector shift thanks to the red-detuned beam (from $0.19$ to $0.05$~mK). Residual splitting is caused by the blue-detuned beam.}
	\label{fig:microtoroid}
\end{figure}

Following Refs.~\cite{Alton2011, Stern2011} we compute with \nanotrappy{} the trapping potential for cesium atoms near a SiO$_2$ microtoroid with a 12 $\mu$m outer major radius $R_{\mathrm{ext}}$, and a 1.5 $\mu$m minor radius $r$. Figure~\ref{fig:microtoroid}(a) shows this structure and the transverse shape of the red-detuned mode used for trapping. The blue-detuned one is not displayed as it is composed of a single lobe in this plane. Modes of the electric field in a microtoroid can be described by their azimuthal number $m$, corresponding to the number of zeros of the field in one turn, and their number $p$, counting the number of lobes in the transverse plane. We choose $m=119$ and $p=0$ for the blue mode and $m=106$ and $p=1$ for the red-detuned beam. The latter has a faster decay in the vertical direction than the blue-detuned one, preventing the atoms from approaching the surface out of the symmetry plane \cite{Vernooy1997, Alton2013}.

Figure~\ref{fig:microtoroid}(b) shows the simulation of a two-color dipole trap in a section of the microtoroid with beams red- and blue-detuned from the cesium D$_2$ line at 852~nm. Lasers with powers $\sim 50$~mW give a trap depth of about 1.5~mK.
Using only one beam of each color produces a strong vector shift at the position of the atoms, manifested by a inhomogeneous broadening of around 0.2~mK (4~MHz) at the trap minimum. Adding a counterpropagating red beam reduces this effect by a factor 4, as shown in Figs. \ref{fig:microtoroid}(c) and (d).
As for the previous examples, the numbers are in very good agreement with the literature. This validates the accuracy of the package and makes it an efficient tool to study optical trapping near nanostructures.  \\

\section{Conclusion} \label{Conclusion}

We have developed a package for simulating optical dipole trap for alkali atoms with an emphasis on trapping atoms close to surfaces. The package can efficiently and accurately calculate 3D trapping potentials, incorporating the scalar, vector and tensor light shifts, for all the Zeeman sublevels of the specified states. It also provides the relevant trap parameters, for example the position of the trap minimum, trap depth and the trap frequencies in all three directions. We provided three example of atom trapping near nanophotonic structures and demonstrated thereby the accuracy of the calculation by comparing our results with published literature.

The scope of application of the \nanotrappy{} package is broad as it can be used to simulate optical dipole traps for any given intensity distribution of the trapping field.
This makes the package appealing to a larger atomic physics community. In addition, the capability of calculating the shifts of the Zeeman levels in a given light field, can be used for estimating dephasing and fidelity of a quantum operation and is useful to the atom-based quantum information community.

\acknowledgments
The authors acknowledge fruitful discussions with Malik Kemiche and Nikos Fayard.
This work was supported by the French National Research Agency NanoStrong Project (ANR-18-CE47-0008), and the R\'{e}gion \^{I}le-de-France (DIM SIRTEQ). This project has also received funding from the European Union\textquotesingle s Horizon 2020 research and innovation programme under Grant Agreement No. 899275 (DAALI project). A.U. was supported by the European Union (Marie Curie Fellowship SinglePass 101030421).


\begin{thebibliography}{68}
\bibitem [{For()}]{Formatting}%
  \BibitemOpen
  \href@noop {} {}\bibinfo {note} {Formatting of the fields should follow a
  predefined convention, see~\cite{Nanotrappyweb}.}\BibitemShut {Stop}%
\end{thebibliography}
\end{document}